\documentclass[runningheads,fleqn]{cl2emult}

\usepackage{makeidx}  % allows index generation
\usepackage{graphicx} % standard LaTeX graphics tool
\usepackage{subeqnar} % subnumbers individual equations
\usepackage{multicol} % used for the two-column index
\usepackage{cropmark} % cropmarks for pages without
\usepackage{phys}     % flushleft layout of math and captions
\makeindex            % used for the subject index
\begin{document}
\title*{Magnetic Properties of Finite Systems:\protect\newline 
Microcanonical Finite-Size Scaling}
\toctitle{Magnetic Properties of Finite Systems:\protect\newline 
Microcanonical Finite-Size Scaling}
% allows explicit linebreak for the table of content
%
%
\titlerunning{Microcanonical Finite-Size Scaling}
% allows abbreviation of title, if the full title is too long
% to fit in the running head
%
\author{Michael Promberger%\inst{1}
\and Michael Kastner%\inst{1}
\and Alfred H\"uller%\inst{1}}
}
\authorrunning{Michael Promberger et al.}
% if there are more than two authors,
% please abbreviate author list for running head
%
%
\institute{Institut f\"ur Theoretische Physik, Staudtstra{\ss}e 7B3, D--91058 Erlangen, Germany}

\maketitle              % typesets the title of the contribution

\begin{abstract}
In the microcanonical ensemble, suitably defined observables show non-analyticities
and power law behaviour even for finite systems. For these observables, a microcanonical finite-size
scaling theory is established which facilitates an approach to the critical exponents of the infinite 
system.
\end{abstract}

\section{Introduction: microcanonical description of finite systems}

In a microcanonical description of finite systems, the magnetic equation of 
state and the zero-field isothermal magnetic susceptibility are defined in terms 
of derivatives of the microcanonical entropy $s({\scriptstyle \varepsilon,m,L^{-1}})$ (cf.~\cite{KPH98},\cite{FN1}):\\
\begin{equation}
\label{mmic}
m_{\scriptscriptstyle {h=0}} ( {\scriptstyle \varepsilon,L^{-1}} ) \quad  \Longleftarrow  \quad
            \max_{m\ge0}\left\{ s({\scriptstyle \varepsilon,m,L^{-1}} ) \right\}\,\,=\,\,
            s\left( {\scriptstyle \varepsilon,m_{\scriptscriptstyle {h=0}} (\varepsilon,L^{-1}),L^{-1}} \right)
\end{equation}
\begin{eqnarray}
\label{chimic}
\lefteqn{
\chi_{\scriptscriptstyle T;{h=0}}({\scriptstyle \varepsilon , L^{-1} } )
\,\, := \,\, \left. \frac{\partial m}{\partial h} \right|_{T;{h=0}}( {\scriptstyle \varepsilon,L^{-1}} )\,\,=}
                     \\
         & & 
                    =\,\, \left\{\left[ \frac{\partial s}{\partial \varepsilon}
                    \left[ \left( \frac{\partial^2 s}{\partial \varepsilon
                    \partial m}\right)^2\left/
                    \frac{\partial^2 s}{\partial \varepsilon^2}\right.
                    -\frac{\partial^2s}{\partial m^2}
                    \right]^{-1}\right]( {\scriptstyle \varepsilon,m,L^{-1}} )\right\}_{m=m_{\scriptscriptstyle {h=0}} (\varepsilon,L^{-1})} 
\nonumber
\end{eqnarray}
Here, $\varepsilon$ denotes the specific enthalpy (cf.~\cite{FN0}), $h$ the magnetic field, $m$ the specific magnetization, 
$L$ the linear system size and $d$ the spatial dimension. For notational convenience, the zero field limit is denoted by $h=0$.
By definition, the microcanonical entropy is the logarithm of the microcanonical partition function (density of states) $\Omega$:
\begin{equation}
s({\scriptstyle \varepsilon,m,L^{-1}} ) \,=\, L^{-d}\,\ln \Omega ({\scriptstyle \varepsilon,m,L^{-1}} )\;\;\;. \label{smic}
\end{equation}
In Fig.~1, the zero-field magnetization (Fig.~1a) and the zero-field magnetic susceptibility (Fig.~1b) are plotted 
for the example of the $3d$-Ising system for various system sizes. Both observables show a behaviour reminiscent 
of the critical behaviour of the infinite system and therefore suggest the introduction of 
finite system exponents $\beta_{\varepsilon,L}$ and $\gamma_{\varepsilon,L}$ (describing the power law behaviour
with respect to the base $|\tilde{\varepsilon} |$ defined below; cf.~Ref.~13 of \cite{KPH98}).\\
The transition energies $\varepsilon_T(L)$, at which the non-analyticities occur, are obviously
dependent on system size.\\
In Figs.~1c/d, magnetizations and susceptibilities are plotted as functions of the reduced enthalpy 
$\tilde{\varepsilon}(L) := (\varepsilon-\varepsilon_T(L))/|\varepsilon_T(L)|$. 
%
%%%%%%%%%%%%%%%%%%%%%%%%%%%%%%%%%%%%%%%%%%%%%%%%%%%%%%%%%%%%%%%%%%%%%%%%%%%%%%%%%%%%%%%%%%%%%%%%
%\input{multibild}
\begin{center}
\begin{figure}[h]

% GNUPLOT: LaTeX picture with Postscript
\setlength{\unitlength}{0.1bp}
%\begin{picture}(1800,1511)(300,0)
%\begin{picture}(1800,1411)(300,0)
\begin{picture}(1800,1611)(300,0)
\put(673,380){\makebox(0,0)[r]{$\scriptstyle L=18$}}
\put(673,480){\makebox(0,0)[r]{$\scriptstyle L=16$}}
\put(673,580){\makebox(0,0)[r]{$\scriptstyle L=14$}}
\put(673,680){\makebox(0,0)[r]{$\scriptstyle L=12$}}
\put(673,780){\makebox(0,0)[r]{$\scriptstyle L=10$}}
\put(1050,1411){\makebox(0,0){{\bf a)} $\qquad m_{\scriptscriptstyle {h=0}}({\scriptstyle \varepsilon,L^{-1}})\,\,$ {\em vs.} $\varepsilon$\rule{19mm}{0mm}}}
\put(1750,150){\makebox(0,0){-0.8}}
\put(1283,150){\makebox(0,0){-0.9}}
\put(817,150){\makebox(0,0){-1}}
\put(350,150){\makebox(0,0){-1.1}}
\put(300,1044){\makebox(0,0)[r]{0.3}}
\put(300,780){\makebox(0,0)[r]{0.2}}
\put(300,515){\makebox(0,0)[r]{0.1}}
\put(300,250){\makebox(0,0)[r]{0}}
\end{picture}
%\begin{picture}(1800,1511)(200,0)
%\begin{picture}(1800,1211)(200,0)
\begin{picture}(1800,1511)(200,0)
\put(1565,726){\makebox(0,0)[r]{$\scriptstyle L=18$}}
\put(1565,826){\makebox(0,0)[r]{$\scriptstyle L=16$}}
\put(1565,926){\makebox(0,0)[r]{$\scriptstyle L=14$}}
\put(1565,1026){\makebox(0,0)[r]{$\scriptstyle L=12$}}
\put(1565,1126){\makebox(0,0)[r]{$\scriptstyle L=10$}}
\put(1075,1411){\makebox(0,0){{\bf b)} $\quad \left[ \frac{\partial^2 s}{\partial m^2}\right]^{-1}({\scriptstyle \varepsilon,m_{\scriptscriptstyle {h=0}},L^{-1}})\,\,$ {\em vs.} $\varepsilon$\rule{7mm}{0mm}}}
\put(1750,150){\makebox(0,0){-0.7}}
\put(1300,150){\makebox(0,0){-0.8}}
\put(850,150){\makebox(0,0){-0.9}}
\put(400,150){\makebox(0,0){-1}}
\put(350,1059){\makebox(0,0)[r]{1200}}
\put(350,789){\makebox(0,0)[r]{800}}
\put(350,520){\makebox(0,0)[r]{400}}
\put(350,250){\makebox(0,0)[r]{0}}
\end{picture}
\begin{picture}(1800,1511)(300,0)
\put(712,370){\makebox(0,0)[r]{$\scriptstyle L=18$}}
\put(712,470){\makebox(0,0)[r]{$\scriptstyle L=16$}}
\put(712,570){\makebox(0,0)[r]{$\scriptstyle L=14$}}
\put(712,670){\makebox(0,0)[r]{$\scriptstyle L=12$}}
\put(712,770){\makebox(0,0)[r]{$\scriptstyle L=10$}}
\put(1050,1411){\makebox(0,0){{\bf c)} $\quad A({\scriptstyle L^{-1}})\,\,\,m_{\scriptscriptstyle {h=0}}({\scriptstyle \tilde{\varepsilon}
,L^{-1}})\,\,$ {\em vs.} $\tilde{\varepsilon}
$\rule{10mm}{0mm}}}
\put(1585,150){\makebox(0,0){0}}
\put(1256,150){\makebox(0,0){-0.04}}
\put(926,150){\makebox(0,0){-0.08}}
\put(597,150){\makebox(0,0){-0.12}}
%\put(300,1117){\makebox(0,0)[r]{0.3}}
%\put(300,828){\makebox(0,0)[r]{0.2}}
%\put(300,539){\makebox(0,0)[r]{0.1}}
%\put(300,250){\makebox(0,0)[r]{0}}
\end{picture}
\begin{picture}(1800,1511)(200,0)
\put(1160,760){\makebox(0,0)[r]{$\scriptstyle L=18$}}
\put(1160,860){\makebox(0,0)[r]{$\scriptstyle L=16$}}
\put(1160,960){\makebox(0,0)[r]{$\scriptstyle L=14$}}
\put(1160,1060){\makebox(0,0)[r]{$\scriptstyle L=12$}}
\put(1160,1160){\makebox(0,0)[r]{$\scriptstyle L=10$}}
\put(1075,1411){\makebox(0,0){{\bf d)} $\,\, B({\scriptstyle L^{-1}})\,\,\, \left[ \frac{\partial^2 s}{\partial m^2} \right] ({\scriptstyle \tilde{\varepsilon}
,m_{\scriptscriptstyle {h=0}},L^{-1}})\,\,$ {\em vs.} $\tilde{\varepsilon}
$\rule{4mm}{0mm}}}
\put(1750,150){\makebox(0,0){0.1}}
\put(1413,150){\makebox(0,0){0.05}}
\put(1075,150){\makebox(0,0){0}}
\put(738,150){\makebox(0,0){-0.05}}
\put(400,150){\makebox(0,0){-0.1}}
%\put(350,1261){\makebox(0,0)[r]{0.1}}
%\put(350,1059){\makebox(0,0)[r]{0.08}}
%\put(350,857){\makebox(0,0)[r]{0.06}}
%\put(350,654){\makebox(0,0)[r]{0.04}}
%\put(350,452){\makebox(0,0)[r]{0.02}}
%\put(350,250){\makebox(0,0)[r]{0}}
\end{picture}
% \vspace{-8mm}
% \vspace{-8mm}
%
% \end{center}
%
%\centerline{\parbox{6.5in}{\caption[%
{\caption[%
]{%
\label{multib}
\small{%
Microcanonical magnetizations and "susceptibilities" of finite 3$d$-Ising systems
(Figs.~a/b). For numerical convenience, $\left[ (\partial^2 s)/(\partial m^2)\right]^{-1}$
is shown instead of $\chi_{\scriptscriptstyle T;{h=0}}$. Note that both quantities show the same
% MFSS
microcanonical finite-size scaling behaviour
and -- in the thermodynamic limit -- the same critical behaviour.
Multiplication of $m_{\scriptscriptstyle {h=0}}$ and $(\partial^2 s)/(\partial m^2)$ by suitable scaling factors
$A({\scriptstyle L^{-1}})$ and $B({\scriptstyle L^{-1}})$ and plotting the thus obtained results
against $\tilde{\varepsilon}$ yields data collapse (Figs.~c/d). Whereas Figs.~a/b suggest the introduction of finite system
exponents, Figs.~c/d give rise to the assumption that the thus defined exponents
show no system size dependence.
}
}}

\end{figure}
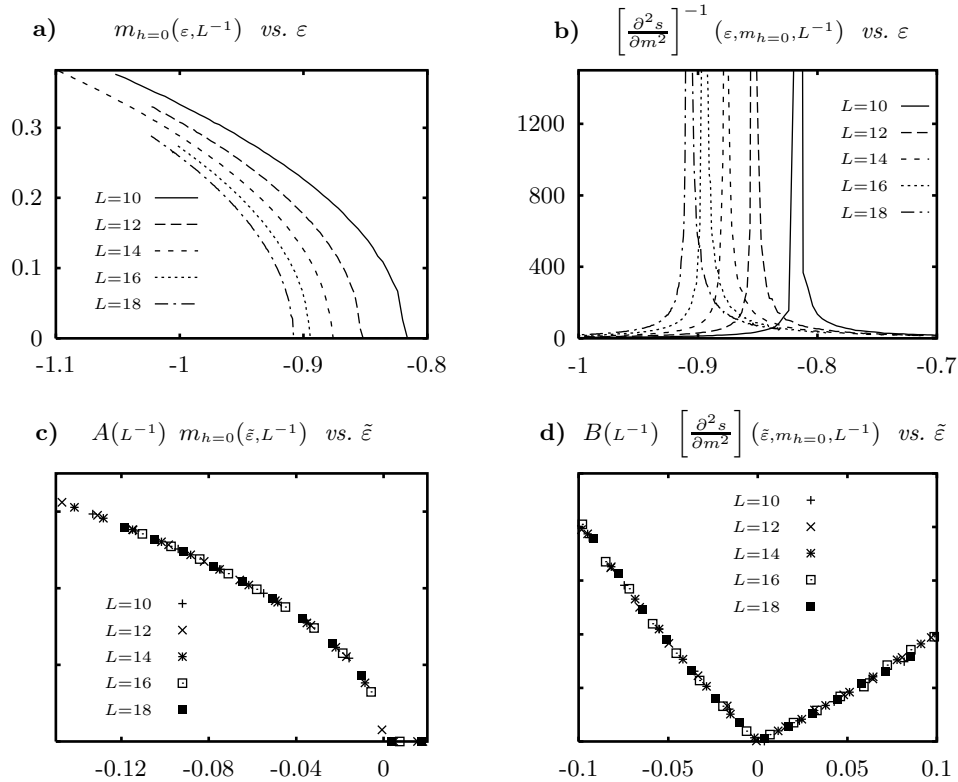
\end{center}
\vspace{-8mm}
%%%%%%%%%%%%%%%%%%%%%%%%%%%%%%%%%%%%%%%%%%%%%%%%%%%%%%%%%%%%%%%%%%%%%%%%%%%%%%%%%%%%%%%%%%%%%%%%
Data collapse is achieved by suitable scaling of the amplitudes of $m_{\scriptscriptstyle {h=0}}$ and $\chi_{\scriptscriptstyle T;{h=0}}$.
Therefore, it seems to be evident that the finite system critical exponents are independent of the system size
(i.e.~take on the same value $\forall L^{-1} \ne 0$), but differ from the values expected in the thermodynamic limit ({\bf TDL}),
$\beta_{\varepsilon,\infty} \approx 0.37$, $\gamma_{\varepsilon,\infty}\approx 1.38$, whereas from numerical data we obtain
$\beta_{\varepsilon,L}\approx 0.5$, $\gamma_{\varepsilon,L}\approx 1$ for all $L$ considered.
\noindent In this paper, a microcanonical finite size scaling ({\bf MFSS}) theory is developed, taking into account
the following constraints (partly justified above):\\
\vspace{-.6cm}

\newcounter{con}
\begin{list}{(\roman{con})}{
\usecounter{con}
\setlength{\leftmargin}{.8cm}
\setlength{\rightmargin}{0cm}
\setlength{\labelsep}{0.25cm}
\setlength{\itemindent}{0cm}
\setlength{\topsep}{2ex}
}
\item consistence with the canonical finite size scaling ({\bf CFSS}) theory\\[-2ex]%[-4.2ex]
\item power law behaviour of the finite system magnetization and susceptibility\\[-2ex]%[-4.2ex]
\item finite system critical exponents which do not depend on the system size and which are not necessarily identical to those of the infinite system\\[-2ex]%[-4.2ex]
\end{list}

\section{Microcanonical finite-size scaling (MFSS)}

In the vicinity of a critical point ($t, h, \varepsilon^*, m \approx 0$), the thermodynamic potentials can be divided
into a singular part (superscript$\phantom{a}^*$) which describes the non-analytic 
behaviour, and a regular part 
(subscript $\phantom{a}_{reg}$). For the % Gibbs 
free energy density and the specific entropy this reads:
\begin{equation}
\label{free_energy}
\hspace{-5mm}
g(t,h) \, =\,  g^*(t,h) \,+\, g_{reg}(t,h) \quad , \quad {\scriptstyle t:=\frac{T-T_c}{T_c} } \quad , \quad
\begin{array}{l} {\scriptstyle T_c}\, \mbox{\scriptsize critical temperature} \\
              {\scriptstyle h}\,\, \mbox{\scriptsize magnetic field}
\end{array}
\end{equation}
\begin{equation}
\label{entropy}
\hspace{-5mm}
s(\varepsilon^*,m) \,=\, s^*(\varepsilon^*,m) \,+\, s_{reg}(\varepsilon^*,m) \quad , \quad 
{\scriptstyle \varepsilon^* := \frac{\varepsilon-\varepsilon_c}{|\varepsilon_c|} } \quad , \quad
{\scriptstyle \varepsilon_c} \,\,\mbox{\scriptsize critical enthalpy}
\end{equation}
It can be shown (see e.g.~\cite{CJL76} and references therein) that in the TDL
the singular parts of the various potentials are homogeneous functions (cf.~Eqs.~(6),(7) in Fig.~2) and 
all critical exponents can be expressed in terms of the degrees of homogeneity
($a_\varepsilon=(1-\alpha)/(2-\alpha)$, $a_m=1/(\delta+1)$, $a_t=1-a_\varepsilon$, $a_h=1-a_m$).
\setcounter{equation}{9}
The equivalence of ensembles is valid only in the TDL. Hence, the thermodynamic potentials of finite systems have to be
classified as \underline{c}anonical or \underline{m}i\-cro\-ca\-no\-ni\-cal quantities (cf.~\cite{KPH98},\cite{KPHpub}).
Starting point for the CFSS is the so-called ``CFSS assumption'' \cite{FB72}: % (cf.~Eq (8)):
For large but finite systems, $g_c^*(t,h,L^{-1})$ is a homogeneous function in accordance with Eq.~(8). \\
We have shown elsewhere \cite{KPHpub} that the MFSS assumption (9) entails (8), whereas proof of the reverse is more difficult.
Note that Eq.~(9) even accounts for the possibility of $s_m^*$ showing no system size dependence at all;
nevertheless, this case results in an $L$-dependence of $g_c^*$ and in CFSS.
In the rest of this paper, we will discuss the consequences of the MFSS assumption
which states that:
$$
\hspace{-10mm}
\mbox{\bf MFSS assumption}: \qquad
\begin{array}{c}
s_m^*({\scriptstyle \varepsilon^*,m,L^{-1}})\mbox{ is a homogeneous function}\\[1ex]
\mbox{of its arguments (cf.~Fig.~2, Eq.~(9))}
\end{array}
$$

\def\ss{\scriptstyle}
\def\sss{\scriptscriptstyle}
\begin{figure}[ht]
%\begin{figure}
\begin{center}
\unitlength1cm
% \begin{picture}(15,8)(2.3,0)
%\begin{picture}(15,6.5)(2.3,0)
\begin{picture}(15,7)(2.3,0)
\put(1.8,5.25){\framebox(4,.7){}}
\put(1.8,3.25){\framebox(4,.7){}}
\put(1.8,.3){\framebox(4,.7){}}
\put(10,5.25){\framebox(4.6,.7){}}
\put(10,3.25){\framebox(4.6,.7){}}
\put(10,.3){\framebox(4.6,.7){}}
\thicklines
\put(1.8,6.3){$ \ss (\mathbf{6}) \,\, g^*(t,h)\,\,=\,\,\lambda^{-1} g^*(\lambda^{a_t}t,\lambda^{a_h}h)  $}
\put(2.5,5.5){Homogeneity of $g^*$}
\put(3.7,4){\vector(0,1){1.2}}
\put(2.7,4.5){$\sss L \to \, \infty$}
\put(2.5,3.5){CFSS assumption}
\put(1.8,2.8){$\ss (\mathbf{8}) \qquad g_c^*(t,h,L^{-1})\,\,=$}
\put(1.8,2.3){$\ss \,\,\quad \lambda^{-1} g_c^*(\lambda^{a_t}t,\lambda^{a_h}h,\lambda^{1/d}L^{-1})$}
\put(10,6.3){$\ss (\mathbf{7})\,\, s^*(\varepsilon^*,m)\,\,=\,\, \lambda^{-1} s^*(\lambda^{a_\varepsilon}\varepsilon^*, \lambda^{a_m}m)$}
\put(10.8,5.5){Homogeneity of $s^*$}
\put(12.2,4){\vector(0,1){1.2}}
\put(12.5,4.5){$\sss L \to \, \infty$}
%
% \put(9.6,2.65){$\ss\left\{ \rule{0cm}{1.3cm} \vphantom{\int \limits^{A}_{A}} \right.$}
\put(10.8,3.5){MFSS assumption}
\put(10,2.8){$\ss (\mathbf{9}) \qquad s_m^*(\varepsilon^*,m,L^{-1})\,\,\,=$}
\put(10,2.3){$\ss  \,\,\quad \lambda^{-1} s_m^*(\lambda^{a_\varepsilon}\varepsilon^*, \lambda^{a_m}m, \lambda^{1/d}L^{-1})$}
\put(12.2,1.9){\vector(0,-1){.8}}
\put(10.2,.6){\scriptsize \underline{M}icrocanonical \underline{F}inite \underline{S}ize \underline{S}caling }
\put(3.7,1.9){\vector(0,-1){.8}}
\put(2,.6){\scriptsize \underline{C}anonical \underline{F}inite \underline{S}ize \underline{S}caling }
\put(7,3.8){$\ss {via~Laplace~T.}$}
\put(9.5,3.6){\vector(-1,0){3}}
\put(7,5.8){$\ss {via~Legendre~T.}$}
\put(6.5,5.6){\vector(1,0){3}}
\put(9.5,5.6){\vector(-1,0){3}}

\end{picture}
\vspace{-5mm}
{\caption[%
]{%
\label{fig1}
\small{%
Homogeneity relations $(\mbox{valid}\; \forall \lambda\, \! > 0)$ for the singular parts of the free energy
and the entropy for finite (Eqs.~(8),(9)) as well as infinite systems (Eqs.~(6),(7)).
In the TDL, a Legendre transform
connects the homogeneity of $g^*$ to the homogeneity of $s^*$ (see Ref.~\cite{S71}),
whereas for finite systems it can be shown that the CFSS assumption is a consequence of the MFSS assumption,
i.e., MFSS is consistent with CFSS \cite{KPHpub}.
}
}}
\end{center}
\end{figure}
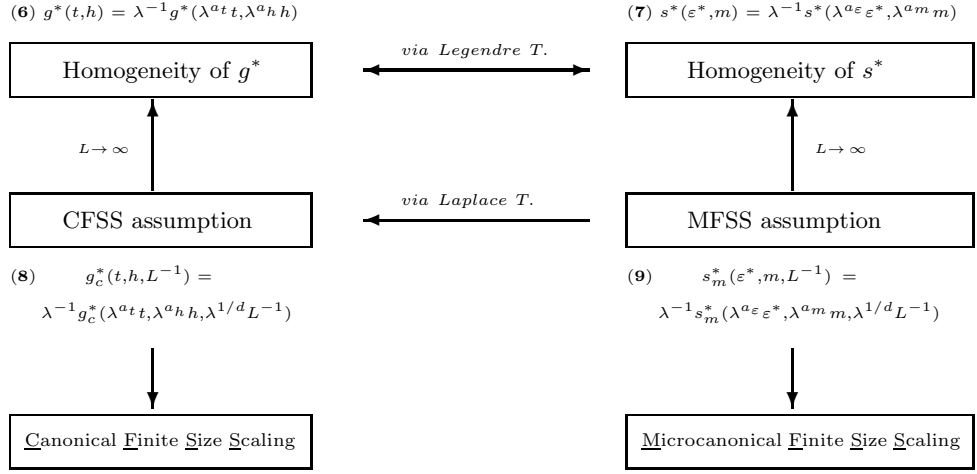
\noindent From the MFSS assumption (9) and Eqs.~(\ref{mmic}) and (\ref{chimic}), the
{\it MFSS relations of the magnetization and the susceptibility} 
are derived easily \cite{FN1}:
\begin{equation}
\label{mfss}
m^*_{\scriptscriptstyle {h=0}} ({\scriptstyle \varepsilon^*,L^{-1}} ) \,\, = \,\, \lambda^{-a_m}m^*_{\scriptscriptstyle {h=0}} ({\scriptstyle \lambda^{a_\varepsilon}\varepsilon^* ,\lambda^{1/d}L^{-1}} )
 \,\,\stackrel{\lambda = L^d}{=} \,\, L^{-da_m} \Phi_{m^*} ({\scriptstyle L^{da_\varepsilon}\varepsilon^*})
\end{equation}
\begin{equation}
\label{chifss}
\hspace{-6mm}
\chi_{\scriptscriptstyle T;{h=0}}^*({\scriptstyle \varepsilon^* , L^{-1} } ) \, = \, \lambda^{1-2a_m} \chi_{\scriptscriptstyle T;{h=0}}^*({\scriptstyle \lambda^{a_\varepsilon}\varepsilon^* , \lambda^{1/d}L^{-1} } ) \,  \stackrel{\lambda = L^d}{=} \,
  L^{d(1-2a_m)}
\Phi_{\chi^*} ({\scriptstyle L^{da_\varepsilon}\varepsilon^*} ) 
\end{equation}

\noindent where the $\Phi_i$ are so-called MFSS functions
which describe the behaviour of the magnetization and the susceptibility of finite systems
in the vicinity of the 
critical point $\varepsilon_c$ of the infinite system.

\noindent {\it MFSS of the transition point:} 
From Figs.~1a/c, it can be deduced that the MFSS function of the magnetization $\Phi_m(x)$ is zero for $L^{da_\varepsilon}\varepsilon^* =: x \ge x_T$, where $x_T$ 
determines the magnetic transition point $\varepsilon_T(L)$ for finite systems:
\begin{equation}
\label{efss}
\hspace{-5mm}
x_T\,=\,L^{da_\varepsilon} \frac{\varepsilon_T(L)-\varepsilon_c}{|\varepsilon_c|} 
\,\,\,\, \Longleftrightarrow \,\,\,\,
\varepsilon_T(L) \,=\, \varepsilon_c (1+\tilde{x}_TL^{-da_\varepsilon})
\, , \,\,
\tilde{x}_T:=\frac{\varepsilon_c}{|\varepsilon_c|}x_T
\end{equation}
\noindent\underline{Remarks:}
a) 
An additional quantity $\tilde{s}_m^*$ can be introduced,
which is defined to be the singular part of the entropy, written in terms of the reduced enthalpy
$\tilde{\varepsilon}(L) := (\varepsilon-\varepsilon_T(L))/|\varepsilon_T(L)|$.  
\newpage
\noindent Using Eq.~(\ref{efss}), the thus defined entropy $\tilde{s}_m^*$ can be
shown to possess the same degrees of homogeneity as $s_m^*$:
\begin{equation}
\label{stild}
\tilde{s}_m^*({\scriptstyle \tilde{\varepsilon},m,L^{-1}} ) \,=\,\lambda^{-1}\tilde{s}_m^*({\scriptstyle \lambda^{a_\varepsilon}\tilde{\varepsilon},\lambda^{a_m}m,\lambda^{1/d}L^{-1}} ) \,\,\, , \,\,\,\mbox{where}
\end{equation}
$$
\tilde{s}_m^*({\scriptstyle \tilde{\varepsilon},m,L^{-1}} ) \,\,:=\,\,s_m^*({\scriptstyle \varepsilon^*=\tilde{\varepsilon}+x_TL^{-da_\varepsilon},m,L^{-1}})\,\,\, .
\phantom{..........................................}
$$
Starting from expression (\ref{stild}), MFSS relations can be derived for the magnetization and the
susceptibility. They illustrate the system size dependence of the behaviour of $m_{\scriptscriptstyle {h=0}}$
and $\chi_{\scriptscriptstyle T;{h=0}}$ in the vicinity of the transition point $\varepsilon_T(L)$ of the finite system:
\begin{equation}
\label{mfsstild}
\tilde{m}_{\scriptscriptstyle {h=0}} ({\scriptstyle \tilde{\varepsilon},L^{-1}} ) \,\, =\,\, 
\lambda^{-a_m}\tilde{m}_{\scriptscriptstyle {h=0}} ({\scriptstyle \lambda^{a_\varepsilon} \tilde{\varepsilon} ,\lambda^{1/d}L^{-1}} )  \,\, \stackrel{\lambda = L^d}{=}
\,\, L^{-da_m} \Phi_{\tilde{m}} ({\scriptstyle L^{da_\varepsilon}\tilde{\varepsilon}} )
\end{equation}
\begin{equation}
\label{chifsstild}
\hspace{-5mm}
\tilde{\chi}_{\scriptscriptstyle T;{h=0}}({\scriptstyle \tilde{\varepsilon} , L^{-1} } ) \,\, = \,\, \lambda^{1-2a_m} \tilde{\chi}_{\scriptscriptstyle T;{h=0}}({\scriptstyle \lambda^{a_\varepsilon}\tilde{\varepsilon} , \lambda^{1/d}L^{-1} } ) \,\, \stackrel{\lambda = L^d}{=} \,\,  L^{d(1-2a_m)}
\Phi_{\tilde{\chi}} ({\scriptstyle L^{da_\varepsilon}\tilde{\varepsilon}} ) 
\end{equation}
b) Numerical data suggest power law behaviour for both the magnetization and the susceptibility of finite systems.
Therefore, the MFSS functions $\Phi_{\tilde{m}}$ and $\Phi_{\tilde{\chi}}$ have to be power laws,
governed by the respective finite system exponents which are \underline{not} determined by the degrees of
homogeneity of $s_m^*$:
\begin{eqnarray}
\Phi_{\tilde{m}} (x) & \propto & (-x)^{\beta_{\varepsilon,L}} \;\;\;\;\;\;\;\;\mbox{for}\;\;x\,\le\,0 \\
\Phi_{\tilde{\chi}} (x) & \propto & |x|^{-\gamma_{\varepsilon,L}} 
\end{eqnarray}
c) $g_c^*$ is analytic for all finite $L$ and shows non-analyticities only in the TDL.
To the best of our knowledge, no proof exists that $s_m^*$ is analytic for finite systems.
Depending on the values of $\beta_{\varepsilon,L}$ and $\gamma_{\varepsilon,L}$, Eqs.~(16) and (17) imply the possibility 
for $s_m^*$ to be either an analytic or a non-analytic function. This means that even for a completely analytic 
entropy, non-analyticities can occur for the microcanonical magnetization and susceptibility.\\
d) Note that it is indeed possible to explicitly construct a function $\tilde{s}_m^*$ %   $(\tilde{\varepsilon},m,L^{-1})$
which complies with the requirements (i)-(iii) stated at the end of the introduction.
This explicit form is quite informative with respect to the "sudden" change of the exponents from
their finite system values towards their values in the TDL, but has to be discussed elsewhere \cite{KPHpub}.

\section{Conclusion}

A MFSS theory has been developed in accordance with the demanded properties (i)-(iii) stated above. 
Amazingly, although the 
scaling laws (7) and (9) comprise identical degrees of homogeneity $a_\varepsilon$ and $a_m$ for the singular parts of the 
entropy of the finite \underline{and} infinite system respectively, they nevertheless can account for power 
law behaviour with different exponents $\beta_{\varepsilon,L}$, $\gamma_{\varepsilon,L}$ for finite and $\beta_{\varepsilon,\infty}$, 
$\gamma_{\varepsilon,\infty}$ for infinite systems. 
As canonical potentials emerge from the microcanonical ones basically by means of a
Laplace transform, it is to be expected that the smoothing properties of this integral transform cause
MFSS to be applicable for smaller systems than CFSS.


\begin{thebibliography}{7}
\bibitem{FN0} In the case of an Ising system, the specific enthalpy $\varepsilon$ is given by the interaction energy: 
$\varepsilon:= -L^{-d} J \sum_{\langle i,j \rangle} S_i S_j$,
i.e., $\varepsilon$ does not contain the field dependent term $h \sum_i S_i$ of the Ising Hamiltonian.
\bibitem{FN1} The zero-field magnetization and the susceptibility can be expressed in terms of $\varepsilon$, $\varepsilon^*$ or $\tilde{\varepsilon}$. The corresponding functions are denoted by $m,\chi$; $m^*\!\!,\chi^*$ or $\tilde{m},\tilde{\chi}$.
\end{thebibliography}
\end{document}